\documentclass[twoside]{article}
\usepackage{qic,epsfig}

\newcommand{\lam}{\lambda}
\newcommand{\meanc}{\overline{C_{\psi}}}

\newcommand{\beq}{\begin{equation}}
\newcommand{\eeq}{\end{equation}}
\newcommand{\beqa}{\begin{eqnarray}}
\newcommand{\eeqa}{\end{eqnarray}}

\newcommand{\da}{\dagger} 
\newcommand{\al}{\alpha} \newcommand{\si}{\sigma}
 \newcommand{\om}{\omega}
 
\newcommand{\la}{\langle}
\newcommand{\ra}{\rangle}

\newcommand{\non}{\nonumber}

\def\jpa#1{{ J.\ Phys.\ A} {\bf#1}}

\def\pra#1{{ Phys.\ Rev. A\/} {\bf#1}}
\def\prb#1{{ Phys.\ Rev. B\/} {\bf#1}}

\def\prl#1{{ Phys.\ Rev.\ Lett.} {\bf#1}}
\def\sci#1{{ Science} {\bf#1}}

\def\annph#1{{ Ann.\ Phys.} {\bf #1}}
\def\pla#1{{ Phys.\ Lett. A\/} {\bf#1}}
\def\rmp#1{{ Rev. \ Mod. \ Phys.} {\bf#1}}
\def\nat#1{{ Nature} {\bf#1}}

\begin{document}
\setlength{\textheight}{8.0truein}    

\runninghead{Title  $\ldots$}
            {Author(s) $\ldots$}

\normalsize\textlineskip
\thispagestyle{empty}
\setcounter{page}{1}

\copyrightheading{0}{0}{2003}{000--000}

\vspace*{0.88truein}

\alphfootnote

\fpage{1}

\centerline{\bf
NON-MARKOVIAN QUANTUM TRAJECTORY UNRAVELLINGS}
\vspace*{0.035truein}
\centerline{\bf OF ENTANGLEMENT}
\vspace*{0.37truein}
\centerline{\footnotesize
BRITTANY CORN}
\vspace*{0.015truein}
\centerline{\footnotesize\it Physics and Engineering Physics, Stevens Institute of Technology}
\baselineskip=10pt
\centerline{\footnotesize\it Hoboken, New Jersey 07030, USA}

\vspace*{10pt}
\centerline{\footnotesize
JUN JING\footnote{Corresponding author: junjing@jlu.edu.cn. }}
\vspace*{0.015truein}
\centerline{\footnotesize\it Institute of Atomic and Molecular Physics, Jilin University}
\baselineskip=10pt
\centerline{\footnotesize\it Changchun 130012, Jilin, China}

\vspace*{10pt}
\centerline{\footnotesize
TING YU\footnote{Corresponding author: ting.yu@stevens.edu. }}
\vspace*{0.015truein}
\centerline{\footnotesize\it Physics and Engineering Physics, Stevens Institute of Technology}
\baselineskip=10pt
\centerline{\footnotesize\it Hoboken, New Jersey 07030, USA}

\vspace*{0.225truein}
\publisher{(received date)}{(revised date)}

\vspace*{0.21truein}

\abstracts{
The fully quantized model of double qubits coupled to a common bath is solved using the quantum state diffusion (QSD) approach in the non-Markovian regime. We have established the explicit time-local non-Markovian QSD equations for the two-qubit dissipative and dephasing models. Diffusive quantum trajectories are applied to the entanglement estimation of two-qubit systems in a non-Markovian regime. In both cases, non-Markovian features of entanglement evolution are revealed through quantum diffusive unravellings in the system state space.
}{}{}

\vspace*{10pt}

\vspace*{3pt}
\communicate{to be filled by the Editorial}

\vspace*{1pt}\textlineskip    
\section{Introduction}

Many important realizations in quantum information, such as quantum computing, quantum communication and quantum cryptography, rely on the control and generation of entanglement \cite{qucomp}. However, the true question arises in how to measure or compute the entanglement of a quantum system in order to effectively use that information in application. For a quantum open system \cite{You1,You2,You3,Rev1,Rev2}, described by a reduced density matrix, most definitions of entanglement pertain to a property of an ensemble, such as entanglement of formation \cite{qucomm}, $E(\rho)$, and concurrence \cite{Wootters}, $C(\rho)$. However for possible applications in quantum information processing, the preparation of and measurement on a desired density matrix of an entangled state, mixed or pure, would be quite cumbersome. A more approachable method for theoretical analysis would be to take advantage of the statistical nature of the quantum system and average over many realizations of a single system in order to infer information about the entanglement of the ensemble. Recently, entanglement unravellings in the Markov regime have been proposed in \cite{ET1,ET2,ET3}. In experiment, single quantum trajectory of a superconducting qubit has been observed by fully control over its environment \cite{nature}. For a general non-Markovian quantum open system \cite{nM1,nM2,nM3,nM4,nM5,nM6}, such a pure state approach is particularly useful for the numerical simulation of the tracking of entanglement information, which is known to be a hard problem due to the lack of a computable entanglement measure and a viable and exact non-Markovian master equation \cite{ET4,ET5,ET6,Yu-Eberly02,ET7,ET8,ET9,Chenetal}.

In this paper, our research serves as a first example of the efficient estimation of entanglement evolution in non-Markovian regimes without using the system density matrix. We derive the exact quantum state diffusion equation for a pure state to estimate the entanglement evolution of a two-qubit system coupled to a bosonic heat bath at zero temperature \cite{Diosi,Strunz99,Yupra99,deVega,Wisemanetal2002,Bassi,QT3,Wang1}. For a general multi-qubit system \cite{JY}, employing quantum trajectories over density matrices becomes enormously advantageous in terms of computational resource. As will be shown in the following, the entanglement computed from trajectories generally provides useful information about the status of the actual entanglement described by the system density matrix. For some initial states, the trajectory entanglement gives an identical estimation of the system entanglement.

Our paper is structured in the following way. In Sec. \ref{model}, we derive the nonlinear non-Markovian QSD equation with a general qubit-environment interaction. Here the notions of mean entanglement trajectories and its estimation of the actual entanglement measure are also outlined. Secs. \ref{diss} and \ref{depha} are dedicated to the dissipative and dephasing models, respectively. The mean entanglement trajectories are compared over various parameters such that the optimal conditions for entanglement are also presented. In Sec. \ref{con}, we conclude the whole paper. The rigorous derivations is left to the Appendices.

\section{Non-Markovian QSD Equation for Two-Qubit Systems}\label{model}

We present a fully quantized model of two uncoupled qubits with respective transition frequencies $\omega_A$ and $\omega_B$ that are coupled to a common zero-temperature heat bath via the interaction Hamiltonian,
\begin{equation}
H_{\rm int}=\sum_{\lam}[g^*_{\lambda} (L_A+\kappa L_B) a^{\dag}_{\lambda} +g_{\lambda}(L^{\dag}_A+\kappa L_B^\dag) a_{\lambda}],
\end{equation}
where $L_A$ and $L_B$ are Lindblad operators describing the interaction of the qubits A and B with the heat bath, respectively, and $\kappa$ $(0\leq \kappa \leq 1)$ is the control parameter that describes the ratio of the qubits' coupling strengths. The formal linear QSD equation \cite{Diosi} describing the dynamics of the quantum state of the qubits, $\psi_t=\psi_t(x^*)$, is given by
\begin{equation}\label{QSD}
\frac{\partial \psi_t}{\partial t}
= - i H_{\rm sys} \psi_t+\mathcal{L} x^*_t\psi_t -
\mathcal{L}^{\dag}\int^t_0 ds M[x_tx^*_s]
\frac{\delta \psi_t}{\delta x^*_s},
\end{equation}
where the system Hamiltonian is $H_{\rm sys}=\frac{\om_A}{2}\si_z^A+\frac{\om_B}{2}\si_z^B$ and the Lindblad operator is $\mathcal{L}=L_A + \kappa L_B$. The Gaussian process $x^*_t$
satisfies $M[x^*_t]=0$, $M[x^*_tx^*_s]=0$ and the bath correlation function
\begin{equation}\label{Mxts}
M[x_tx^*_s]=\sum_\lambda |g_\lambda|^2e^{-i\omega_\lambda(t-s)}\equiv\al(t,s),
\end{equation}
where $M[\cdot]$ denotes the statistical mean over the noise. The solution to the QSD equation (\ref{QSD}), $\psi_t$, recovers the reduced density matrix of the qubit system: $\rho_t=M[|\psi_t\ra\la \psi_t|]$. Central to the application of the QSD equation is to replace the functional derivative with a time-local operator, termed as the O-operator, such that
\begin{equation}\label{fd}
\frac{\delta \psi_t}{\delta x^*_s}=\hat{\mathcal{O}}(t,s,x^*)\psi_t,
\end{equation}
with initial condition $\hat{\mathcal{O}}(t=s,s,x^*)=\mathcal{L}$. In principle, the existence of the O-operator can be seen from the stochastic propagator, $|\psi_t(x^*)\ra=G(t,x^*)|\psi_0\ra$ (See Appendix A), but in practice it is difficult to find the explicit O-operator. For the specific two-qubit model presented in this paper, an exact equation for the O-operator is derived upon satisfying the consistency condition \cite{Diosi}:
\begin{equation}\label{conscond}
\frac{\delta}{\delta x^*_s}\left(\frac{\partial\psi_t}{\partial t}\right)=
 \frac{\partial}{\partial t}\left(\frac{\delta\psi_t}{\delta x^*_s}\right).
\end{equation}
With the initial condition, it ensures that $\psi_t$ is a single-valued function and thus establishes a solvable QSD equation.

We explore the non-Markovian regime by modeling the bath correlation function as an Ornstein-Uhlenbeck process such that $\alpha(t,s)=\frac{\gamma}{2}e^{-\gamma|t-s|}$. This continuous random process drifts toward a stationary long-term mean and is useful for viewing various memory effects via the parameter $\gamma$, which describes the rate at which noise that is progressing in time $t$ becomes less and less correlated to its value at a particular past time $s$. As $\gamma$ grows very large and the correlation time $\tau_c=\frac{1}{\gamma}$ becomes very short, we will view the transition from non-Markovian to Markovian regimes and find that certain features are lost under the Markov approximation \cite{QuantTraj1,QuantTraj2,Gisin-Percival93,Note}.

We investigate two mechanisms of qubit decoherence: dissipation and pure phase relaxation of the quantum state, both being great challenges to maintaining robust entanglement. In both cases, we have derived the exact time-local O-operators, allowing us to efficiently solve the nonlinear QSD equation \cite{Diosi}, which robustly preserves the norm of the qubit state vector throughout every trajectory, a facet not exhibited by the linear equation \cite{Strunz99}. The dynamics of the normalized quantum state of the qubits, $\tilde{\psi}_t=\frac{\psi_t}{||\psi_t||}$, is presented as \cite{Diosi}:
\begin{eqnarray} \label{NLQSD}
\frac{d\tilde{\psi}_t}{dt} &=&- i H_{\rm sys} \tilde{\psi}_t +
(\mathcal{L}-\la\mathcal{L}\ra_t)\tilde{x}^*_t\tilde{\psi}_t \non \\
&-& \int^t_0 ds \alpha(t,s)[(\mathcal{L}^{\dag}-\la\mathcal{L}^{\dag}\ra_t)
\hat{\mathcal{O}}(t,s,\tilde{x}^*) \non \\
&-&\la(\mathcal{L}^{\dag}-\la\mathcal{L}^{\dag}\ra_t)\hat{\mathcal{O}}
(t,s,\tilde{x}^*)\ra_t]\tilde{\psi}_t,
\end{eqnarray}
where $\la A\ra_t=\la\tilde{\psi}_t|A|\tilde{\psi}_t\ra$ is the quantum expectation value of operator $A$ and $\tilde{x}^*_t=x^*_t+\int^t_0 ds \alpha^*(t,s) \la\mathcal{L}^{\dag}\ra_s$ is the shifted noise. Solving the stochastic differential equation above for a particular realization of the Gaussian random noise reveals a single unraveling of the quantum system evolution, allowing one to calculate a single entanglement trajectory by $C(\psi_t)=|\la \tilde{\psi}_t| \sigma^A_y\otimes\sigma^B_y|\tilde{\psi}^*_t\ra|$. After a large number of realizations are produced, we take the mean over all concurrence trajectories, $\overline{C_{\psi}}\equiv M[C(\psi_t)] $, and obtain a value that is pertinent to the entanglement of the ensemble system. In this way, we can efficiently compute the approximate entanglement of a quantum open system without invoking the explicit form of the density matrix. The actual entanglement represented by the density matrix can be calculated through concurrence \cite{Wootters}
\begin{equation}
C(\rho)=\max\left\{0,\sqrt{\lam_1}-\sqrt{\lam_2}-\sqrt{\lam_3}-\sqrt{\lam_4}
\right\},
\end{equation}
where $\lam_i\,\, (i=1,2,3,4) $ are the eigenvalues of the matrix $\varrho=\rho(\sigma^A_y\otimes\sigma^B_y)\rho^*(\sigma^A_y\otimes \sigma^B_y)$ in descending order. Upon direct comparison, it is clear that $\meanc$ must be greater than or equivalent to the true entanglement $C(\rho)$ due to the concavity of the concurrence calculation \cite{Wootters,ET7}. Therefore, $\meanc$ can be used as an upper bound of the actual entanglement, such that if $\meanc\approx0$ then $C(\rho) \approx0$. In fact, as shown below, $\meanc$ provides an perfect estimation of the actual entanglement for some initial states. Above all, the calculation of $\meanc$ is much simpler than that of $C(\rho)$, especially for systems consisting of a large number of qubits or qudits, where a good entanglement definition of mixed states is not available now. This pronounces $\meanc$ to be a good indicator for the actual behavior of the entanglement and will be explored in the upcoming models.

\section{Dissipative Model}\label{diss}

A dissipative interaction, which causes the quantum state to lose energy as well as coherence, is denoted by the Lindblad operators $L_A=\sigma^A_-$ and $L_B=\sigma^B_-$ such that $\mathcal{L}=\sigma^A_-+\kappa\sigma^B_-$. By the consistency condition of Eq.~(\ref{conscond}), we find the exact operator $\bar{\mathcal{O}}(t, {x}^*)\equiv \int^t_0 ds\al(t,s)\hat{\mathcal{O}}(t,s, {x}^*)$ to be
\begin{eqnarray}\non
\bar{\mathcal{O}}(t, {x}^*)&=&A(t)\sigma^A_-+B(t)\sigma^B_-
+F(t)\sigma^A_z\sigma^B_-+G(t)\sigma^B_z\sigma^A_-\\ \label{tqDO}
&+&i\left[\int^t_0 ds' P(t,s'){x}^*_{s'}\right]\sigma^A_-\sigma^B_-,
\end{eqnarray}
which is valid for an arbitrary bath correlation function. By imposing the Ornstein-Uhlenbeck bath correlation function, we derive (See Appendix B) a set of differential equations for the coefficients of the $\bar{O}$ operator:
\begin{eqnarray}
d_tA(t)&=&-\gamma A(t) +\frac{\gamma}{2}+i\omega_A A(t)+ A^2(t) \non \\  &+& 2\kappa F(t)G(t)+  G^2(t) - \frac{\kappa}{2}iQ(t),\non \\
d_tB(t)&=&-\gamma B(t) +\frac{\gamma\kappa}{2}+i\omega_B B(t)+\kappa B^2(t)\non \\ &+&2F(t)G(t)+ \kappa F^2(t) - \frac{1}{2}iQ(t), \non\\
d_tF(t)&=&-\gamma F(t) +i\omega_B F(t) + F(t)[A(t)+G(t)] \non \\ &+&
B(t)[G(t)-A(t)]+2\kappa B(t)F(t) - \frac{1}{2}iQ(t),\non \\
d_tG(t)&=&-\gamma G(t) +i\omega_A G(t) + \kappa F(t)[A(t)+G(t)]\non \\ &+& \kappa B(t)[G(t)-A(t)]+2A(t)G(t)- \frac{\kappa}{2}iQ(t), \non \\
d_tQ(t)&=&-2\gamma Q(t) + i(\omega_A + \omega_B) Q(t)\non \\ &+& 2[A(t)+\kappa B(t)]Q(t)-i\gamma[F(t)+\kappa G(t)] \label{ABFGQ}
\end{eqnarray}
together with the explicit solution
\begin{eqnarray}\non
P(t,s')&=&-2i[F(s')+\kappa G(s')]\exp\bigg\{\int^t_{s'}ds[-\gamma +
i\omega_A\\ \label{sPts} &+& i\omega_B+2A(s) + 2\kappa B(s)]\bigg\}
\end{eqnarray}
and initial conditions $A(0)=B(0)=F(0)=G(0)=Q(0)=0$. It should be noted that when $\kappa=1$ and $\omega_A=\omega_B$, $A(t) =B(t)$ and  $F(t)=G(t)$, representing a highly symmetrical setup where qubit A and B are interchangeable. Throughout the rest of the paper, we will assume $\omega_A=\omega_B=\omega$.

Knowledge of the exact equations for the O-operator allows us to solve the nonlinear QSD Equation for various unravelings of the time evolution of the qubits initially in the maximally entangled Bell States,
$|\Psi^{\pm}\ra=\frac{1}{\sqrt{2}}\left(|\uparrow\uparrow\ra\pm|
\downarrow\downarrow\ra\right)$ for qubits with correlated spins and
$|\Phi^{\pm}\ra=\frac{1}{\sqrt{2}}\left(|\uparrow\downarrow\ra\pm|
\downarrow\uparrow\ra\right)$ for qubits with anti-correlated spins. In the interest of direct comparison to the exact case, we have derived the $\bar{O}$ operator in the Post-Markov approximation to be
\begin{equation}
\bar{O}_{PM}=[f_0(t)+i\omega f_1(t)]\mathcal{L} - f_2(t)(\sigma^A_z + \kappa^2\sigma^B_z) \mathcal{L},
\end{equation}
where $f_0(t)=\int^t_0 \alpha(t,s)ds $, $f_1(t)=\int^t_0\alpha(t,s)(t-s)ds$, and $f_2(t)=\int^t_0 \int^s_0\alpha(t,s)\alpha(s,u)(t-s)du\hspace{1pt}ds$ \cite{Note}.

\begin{figure}[htbp]
\centering
\includegraphics[width=3.0in]{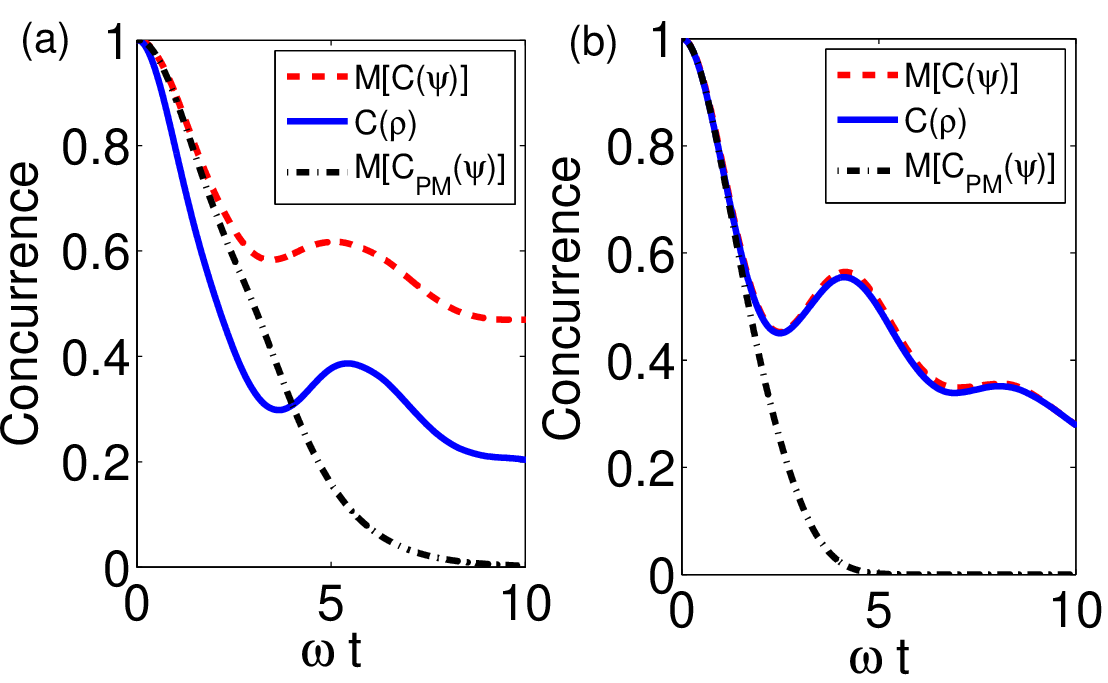}
\fcaption{Dissipative Model: The exact $\meanc$ is compared to ensemble calculations in the non-Markovian regime, $C(\rho)$, and trajectory methods under the Post-Markov approximation, $\overline{C_{PM}(\psi)}$ with $\kappa=1$ and $\gamma=0.3$ for (a) $|\psi_0\ra=|\Psi^{+}\ra$ and (b) $|\psi_0\ra=|\Phi^{+}\ra$.}
\label{DissCSrp}
\end{figure}

\begin{figure}[htbp]
\centering
\includegraphics[width=3in]{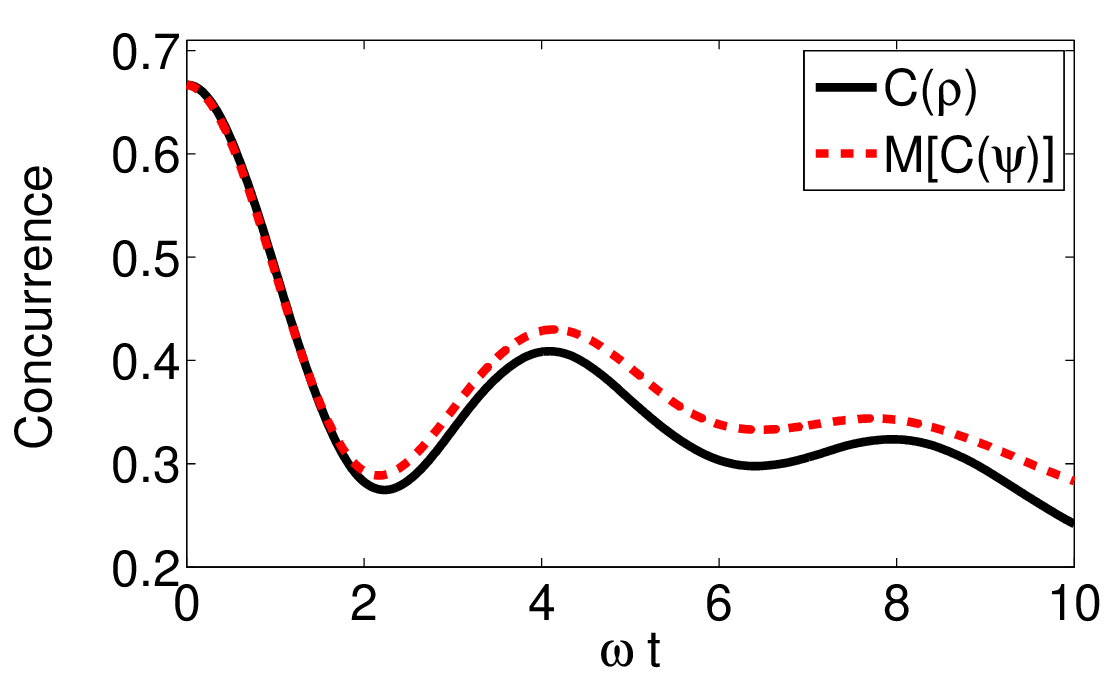}
\fcaption{Dissipative Model: The exact $\meanc$ is compared to ensemble calculations in the non-Markovian regime, $C(\rho)$ with $\kappa=1$ and $\gamma=0.3$ for $|\psi_0\ra=(1/\sqrt{15})(|\uparrow\uparrow\ra+2|\uparrow\downarrow\ra
+3|\downarrow\uparrow\ra+|\downarrow\downarrow\ra)$.}
\label{s1231}
\end{figure}

\begin{figure}[!h]
\centering
\includegraphics[width=3.0in]{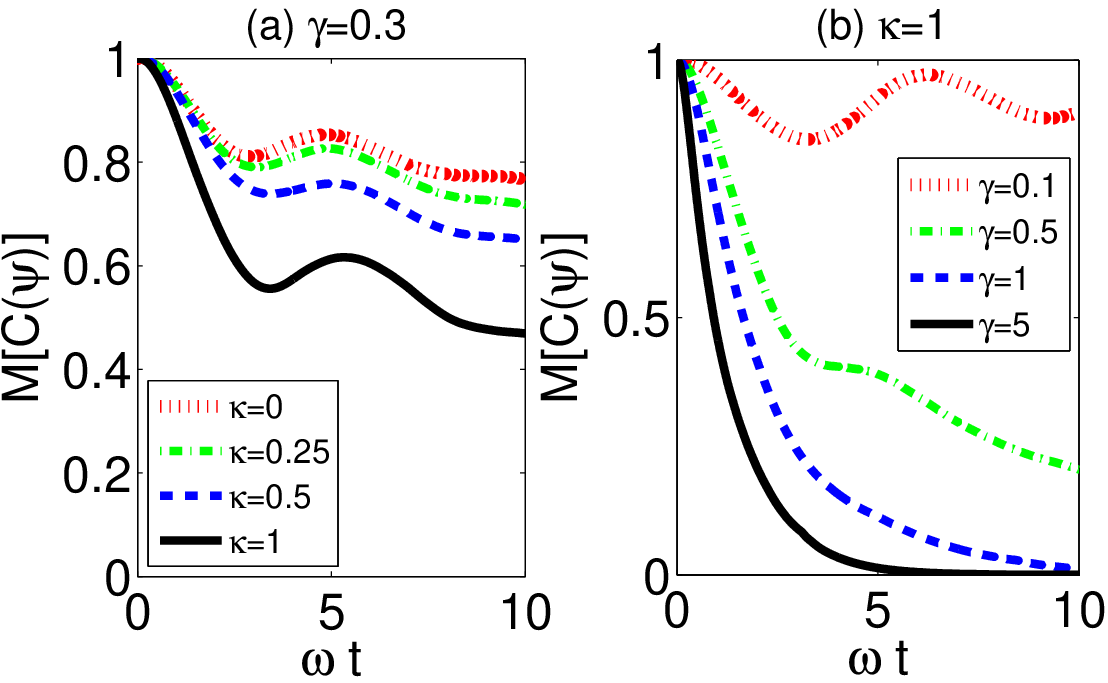}
\fcaption{Dissipative Model: For $|\psi_0\ra=|\Psi^{+}\ra$,
$\overline{C_{\psi}}$  over 1000 realizations is compared over (a) various values of $\kappa$ for fixed $\gamma=0.3$ and (b) various values of $\gamma$ for fixed $\kappa=1.$}
 \label{DissCSkg}
\end{figure}

\begin{figure}[h!]
\centering
 \includegraphics[width=3.0in]{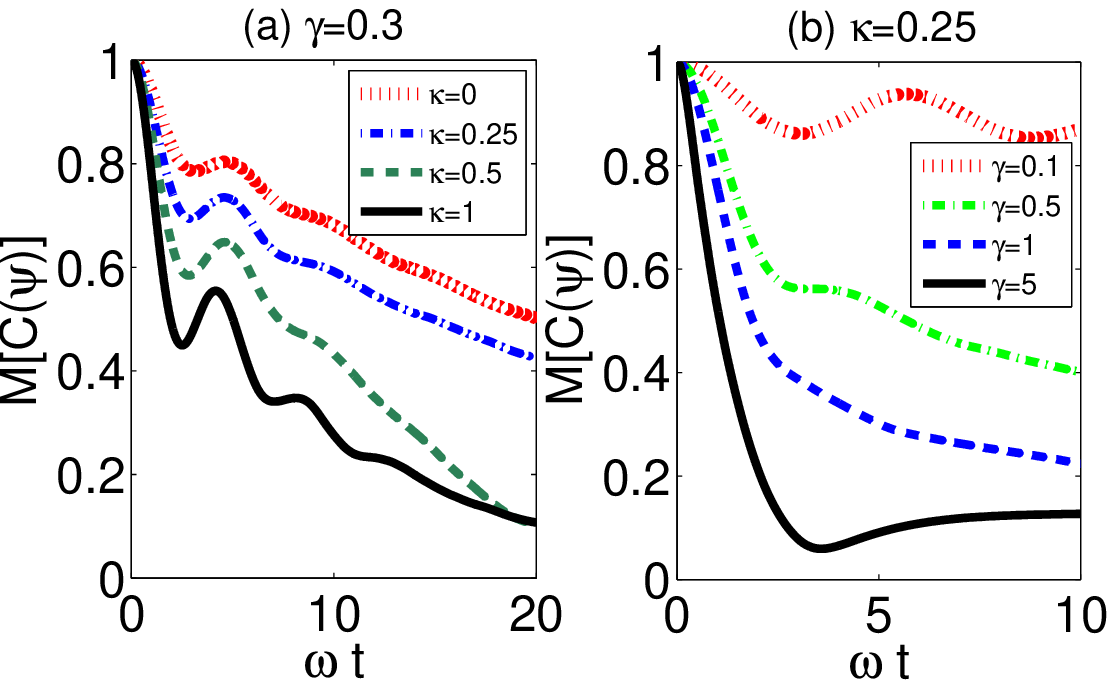}
 \fcaption{Dissipative Model: For $|\psi_0\ra=|\Phi^{+}\ra$,
$\overline{C_{\psi}}$ is compared over (a) various values of $\kappa$ for fixed $\gamma=0.3$ for long times and (b) various values of $\gamma$ for fixed $\kappa=0.25$.}
 \label{DissACSkg}
\end{figure}

The entanglement computed from the exact mean trajectory method, $\meanc$ is compared to that under the Post-Markov approximation, $\overline{C_{PM}(\psi)}$, as well as the entanglement of the ensemble, $C(\rho)$, in Fig.~\ref{DissCSrp}. For both initial states, $|\Psi^{+}\ra$ in (a) and $|\Phi^{+}\ra$ in (b), the actual entanglement $C(\rho)$ displays the repeated revival and decay of entanglement known to the two-qubit model \cite{Braun,Revival,Hu2009}, and solved exactly here and in Ref. \cite{Zhao}. These trends are exhibited by the exact mean entanglement trajectory $\meanc$, whereas in contrast, applying the Post-Markov approximation removes all revival features of the curve. This figure demonstrates the dependence of the theory on the initial qubit state, where $|\Phi^{+}\ra$ provides a much closer approximation than $|\Psi^{+}\ra$. This exactness in the entanglement estimation relies on the initial states. In Fig.~\ref{s1231}, we starts from the state $|\psi_0\ra=(1/\sqrt{15})(|\uparrow\uparrow\ra+2|\uparrow\downarrow\ra
+3|\downarrow\uparrow\ra+|\downarrow\downarrow\ra)$, which could be considered as an extrapolation of the previous two kinds of Bell states. It is shown that during $0\leq\om t<2$, $\meanc$ is perfectly the same as $C(\rho)$. Also it captures all the oscillations during the time evolution afterwards.

Therefore in any case, $\meanc$ acts as an upper-bound for the exact entanglement $C(\rho)$, giving valuable information about the general trends of the entanglement evolution, such as the regeneration of entanglement due to the common bath and memory effect of the environment. Upon taking a closer look at $\meanc$ for various coupling strengths and correlation times in Figs.~\ref{DissCSkg} and~\ref{DissACSkg}, many interesting attributes of this model are revealed and the optimal conditions for entanglement are discussed.

For initial state $|\psi_0\ra=|\Psi^{+}\ra$, Fig.~\ref{DissCSkg}(a) highlights the significant revival feature of the equal couplings case, $\kappa=1$, which shrinks as we decrease the coupling strength of qubit B. However, for this model, the asymmetry of the coupling constants causes the entanglement to decay at a much slower rate and also maintains the qubits in a higher level of entanglement for a significant period of time. This is also due to the memory effects of the non-Markovian environment with $\gamma=0.3$, which generally allows the entanglement to remain non-zero for an extended time. In Fig.~\ref{DissCSkg}(b), the mean entanglement trajectory is compared over various correlation times for the case of qubits with symmetrical coupling, capturing the transition from non-Markovian to Markovian regimes as $\gamma$ becomes large. It is clearly shown that the revival peak of the entanglement grows as we tend toward non-Markovian conditions and eventually oscillates very close to an entangled state that will not decay, allowing one to maintain a highly entangled state over a long period of time when large memory effects are present. The importance of non-Markovian environments becomes apparent in comparison to the Markovian case, $\gamma=5$, where the qubits are plagued with a swift decoherence and have no chance to be re-entangled.

In Fig.~\ref{DissACSkg}(a) we look at long time entanglement evolution from initial state $|\Phi^{+}\ra$ for various values of $\kappa$ and fixed $\gamma=0.3$ where many revival peaks are witnessed. Similar to the previous case of initial state $|\Psi^{+}\ra$, tall revival peaks are displayed for symmetrical couplings, however they come at the expense of a faster disentanglement. Once again, the $\kappa=0$ case reveals a much slower entanglement decay and remains non-zero even for long times. Comparing the effects of memory on the entanglement dynamics, Fig.~\ref{DissACSkg}(b) again demonstrates that a very long correlation time allows the quantum state to remain highly entangled for extended times. An interesting difference for this initial state is that even for fairly large $\gamma=5$, the rebirth of entanglement is still a dominant feature.

\section{Dephasing Model}\label{depha}

\begin{figure}[h!]
 \centering
 \includegraphics[width=3.0in]{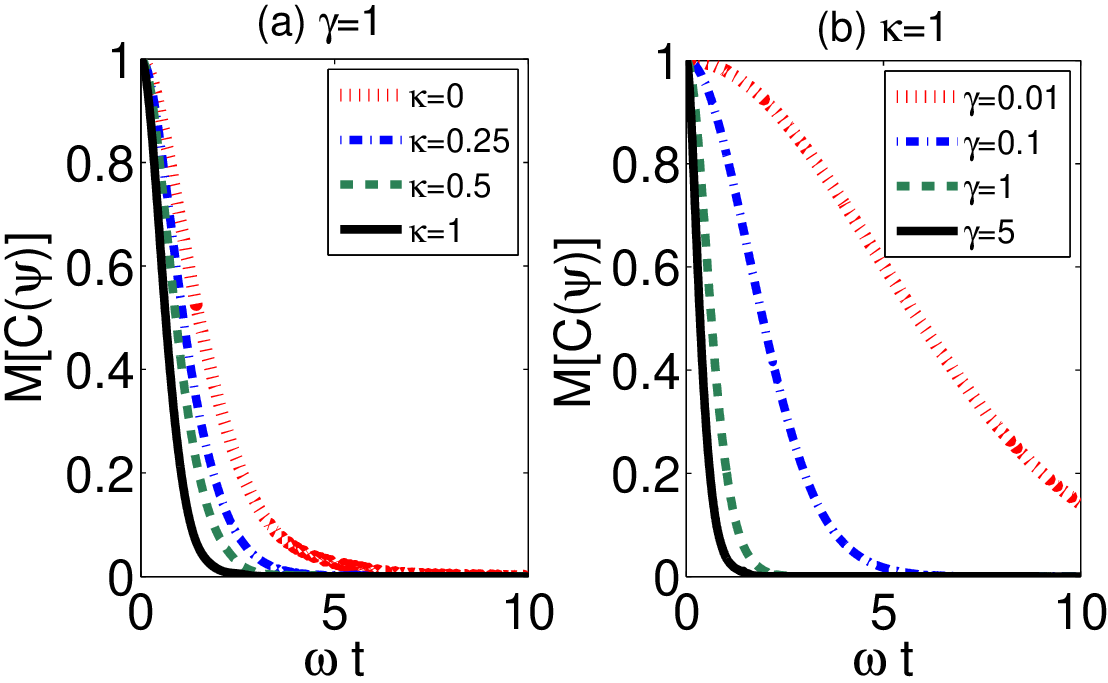}
 \fcaption{Dephasing Model: For $|\psi_0\ra=|\Psi^{\pm}\ra$,
$\overline{C_{\psi}}$ over 1000 realizations is compared over (a) various values of $\kappa$ for fixed $\gamma=1$ and (b) various values of $\gamma$ for fixed $\kappa=1$.}
 \label{DephCSkg}
\end{figure}

\begin{figure}[h!]
 \centering
 \includegraphics[width=3.0in]{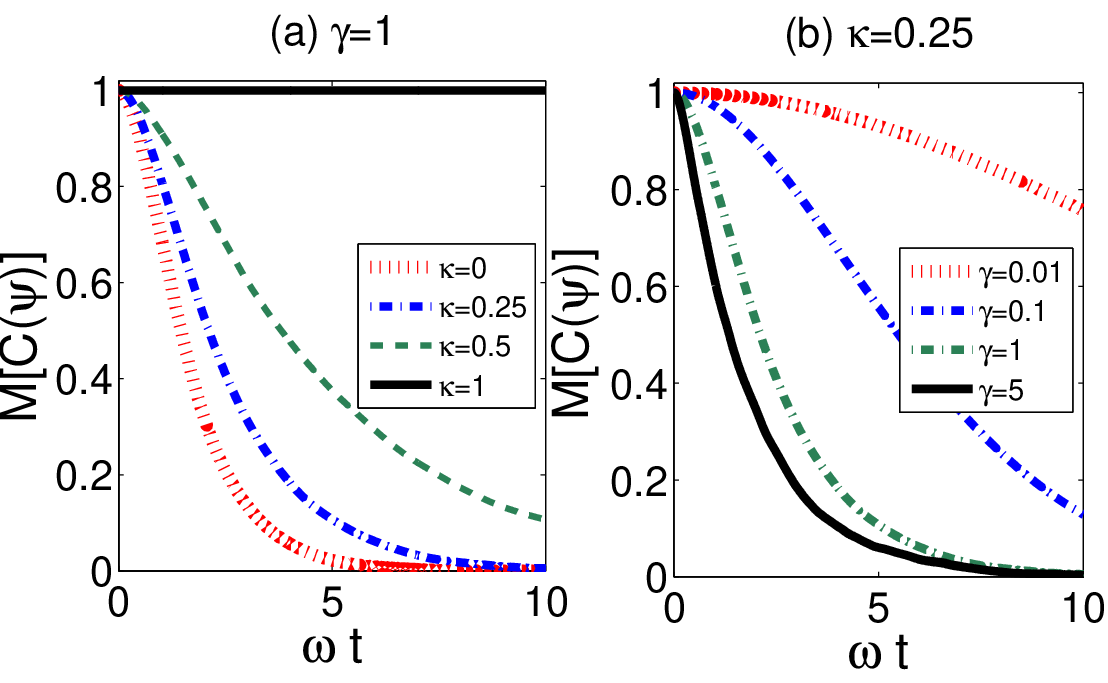}
 \fcaption{Dephasing Model: For $|\psi_0\ra=|\Phi^{\pm}\ra$,
 $\overline{C_{\psi}}$ is compared over (a) various values of $\kappa$ for fixed $\gamma=1$ and (b) various values of $\gamma$ for fixed $\kappa=0.25$. }
\label{DephACSkg}
\end{figure}

As another important case, we consider a dephasing type of interaction, which provides an example of pure decoherence with energy conservation. Described by the two Lindblad operators $L_A=\sigma^A_z$ and $L_B=\sigma^B_z$, the consistency conditions of Eq.~(\ref{conscond}) result in the exact and noise-free O-operator $\hat{\mathcal{O}}(t,s)=\mathcal{L}=\sigma^A_z+\kappa\sigma^B_z$ due to $[H_{\rm sys},\mathcal{L}]=0$ and $\mathcal{L}^\da=\mathcal{L}$. Applying the Ornstein-Uhlenbeck bath correlation function then results in the noise-independent time-local operator $\bar{\mathcal{O}}(t)=\frac{1}{2}(1-e^{\gamma|t|})(\sigma^A_z+\kappa\sigma^B_z)$ that facilitates a solution to the exact non-Markovian QSD equation. The mean entanglement trajectories for the dephasing model are plotted in Figs. (\ref{DephCSkg}) and (\ref{DephACSkg}).

\begin{figure}[h!]
\centering
\includegraphics[width=3in]{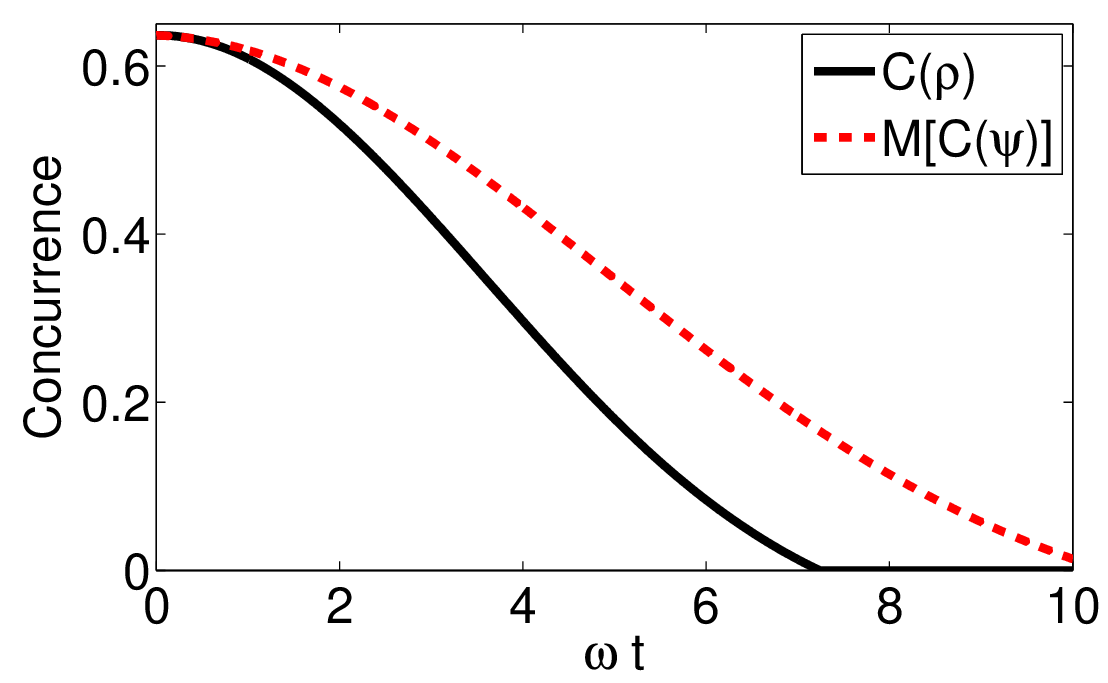}
\fcaption{Dephasing Model: The exact $\meanc$ is compared to ensemble
calculations in the non-Markovian regime, $C(\rho)$ with $\kappa=1$
and $\gamma=0.01$ for
$|\psi_0\ra=(1/\sqrt{22})(2|\uparrow\uparrow\ra+|\uparrow\downarrow\ra
+|\downarrow\uparrow\ra+4|\downarrow\downarrow\ra)$.}
\label{s2114}
\end{figure}

In Fig.~\ref{DephCSkg}(a), the dotted curve, $\kappa=0$, represents the scenario of qubit A interacting with the heat bath while qubit B is a free particle. As is expected of the single-qubit dephasing channel \cite{Yu-Eberly02}, the entanglement of the qubits asymptotically decays to zero. Moreover, as we introduce the interaction of qubit B to the environment through $\kappa\neq0$, the disentanglement rate between the qubits only increases and causes a faster death of entanglement. In Fig.~\ref{DephCSkg}(b), the very non-Markovian case, $\gamma=0.01$, where the memory of the system extends much further into the past, reveals the preservation of high level entanglement for a considerable length of time. In the limit as $\gamma$ approaches $0$ one would expect entanglement to be sustained at the maximum value eternally. As we shorten the memory of the system in the Markov regime, the entanglement curves reveal a steeper and steeper descent toward zero entanglement.

The same analysis was applied to qubits with initial state $|\Phi^{\pm}\ra$ and shown in Fig.~\ref{DephACSkg}. In Fig.~\ref{DephACSkg}(a) we immediately see that, when the coupling constants of the qubits to the heat bath are equal, $\kappa=1$, then the initially entangled state is protected due to the symmetry between the two qubits. When the qubits are not coupled to the modes of the heat bath in exactly the same way, $\kappa\neq 1$, the entanglement will eventually decay to zero. Similar to the dissipative model, Fig.~\ref{DephACSkg}(b) displays the prolonged entanglement of the qubits in the non-Markovian case, $\gamma=0.01$, and the faster disentanglement rate of the Markov approximation, $\gamma=5$.

We also compare $\meanc$ and $C(\rho)$ in the dephasing case. In Fig.~\ref{s2114}, the initial state is chosen beyond $|\Phi^{\pm}\ra$ and $|\Psi^{\pm}\ra$ and the dynamics is computed in a very non-Markovian regime. $\meanc$ also shows a good estimation over the actual entanglement. Since in the dephasing dynamics caused by the Ornstein-Uhlenbeck bath, there is no chance for the entanglement to get revival so that the estimation seems to be not as perfect as that in the dissipation dynamics.

\section{Conclusion}\label{con}

We have shown that the dynamical entanglement of a non-Markovian open system can be efficiently estimated by employing exact quantum diffusive trajectories. In particular, we have shown that the entanglement dynamics of the system are very sensitive to which initial state the qubits evolved from, how the qubits are coupled to the heat bath, and the correlation time of the environment. We emphasized that under the Markov approximation, the entanglement for both sets of Bell states was characterized by fast disentanglement and suppressed revival features; whereas in the non-Markovian regime, large revivals were witnessed and for an extensively long correlation time, the qubits remained nearly maximally entangled for long times. We demonstrate that the optimal conditions for maintaining a high level of entanglement for long periods of time are the symmetric coupling strengths of the qubits to the environment and for the autocorrelation time of the environment to be very long. This again emphasizes the importance of memory effects on the dynamics of a quantum open system. The trajectory estimation provides a good upper-bound (perfect for special initial states) entanglement for the system and it would be more meaningful for a higher dimensional system.

\nonumsection{Acknowledgements}
\noindent
We acknowledge grant support from the NSF PHY-0925174, AFOSR No.~FA9550-12-1-0001, and the NSFC No. 11175110.

\nonumsection{References}

\appendix

\section*{Existence of the O-operator}

In this paper, Eq.~(\ref{QSD}) \cite{Diosi97,Diosi} describes the dynamics of the pure quantum state $\psi_t$ under the influence of the complex stochastic Gaussian process $x^*_t$. The formal linear QSD equation becomes non-local because of the functional derivative with respect to noise, so the time-local non-Markovian QSD equation cannot be derived if the functional derivative cannot be replaced with a linear operator acting on the state vector $\psi_t$. In many physically interesting models \cite{Yupra99,Strunz99,Strunz99-2,Strunz01,Strunz04,Yu04,Jing10}, the functional derivative has been replaced by a linear operator, termed the O-operator, as shown in Eq.~(\ref{fd}). Then Eq.~(\ref{QSD}) turns out to be:
\begin{equation}
\frac{\partial\psi_t}{\partial t}=-iH_{\rm sys}\psi_t+\mathcal{L}x^*_t\psi_t-
\mathcal{L}^{\dag}\int^t_0ds\alpha(t,s)O(t,s,x^*_s)\psi_t,
\end{equation}
where the relation in Eq.~(\ref{Mxts}) has been used. The existence of the O-operator can be seen from the linear propagator $\hat{G}$ for Eq.~(\ref{QSD}), where $\psi_t\equiv\psi_t(x^*)=\hat{G}(t,x^*)\psi_0$ \cite{Feynman}. Consequently,
\begin{eqnarray*}
\frac{\delta\psi_t(x^*)}{\delta x_s^*}=\left[\frac{\delta}{\delta x_s^*}\hat{G}(t,x^*)\right]\psi_0=\left(\frac{\delta}{\delta x_s^*}\hat{G}\right)\hat{G}^{-1}\psi_t(x^*)=\hat{\mathcal{O}}(t,s,x^*)\psi_t(x^*).
\end{eqnarray*}
By the consistency condition in Eq.~(\ref{conscond}), one gets the equation of motion for the O-operator:
\begin{eqnarray}\label{consis2}
\frac{\partial\hat{\mathcal{O}}(t,s,x^*)}{\partial t}=[-iH_{\rm
sys}+\mathcal{L}x^*_t-\mathcal{L}^{\dag}\bar{\mathcal{O}}(t,x^*),
\hat{\mathcal{O}}(t,s,x^*)]-\mathcal{L}^{\dag}\frac{\delta \bar{\mathcal{O}}(t,x^*)}{\delta x^*_s},
\end{eqnarray}
where $\bar{\mathcal{O}}(t,x^*)=\int^t_0ds\alpha(t,s)\hat{\mathcal{O}}(t,s,x^*)$. In general non-Markovian models, it is very challenging to determine the O-operator from Eq.~(\ref{consis2}). At the moment, it was still unclear under what mathematical conditions an exact O-operator can be determined, however it is known that perturbative O-operators can always be obtained \cite{Yupra99}. Moreover, we are able to find the explicit form of the O-operator for this specific two-qubit model. To the best of our knowledge, the exact O-operator has been established only in the following cases: one qubit in a dephasing and dissipative environment \cite{Diosi,Yupra99,Strunz99}, one harmonic oscillator in Brownian motion and dissipative environment \cite{Diosi,Yupra99,Strunz99-2,Strunz01}, one harmonic oscillator in Brownian motion with finite temperature \cite{Strunz04}, a cavity mode in a dissipative environment with zero and finite temperature \cite{Yu04}, one three-level atom in a dissipative environment \cite{Jing10}, multi-level atomic systems \cite{multi}, and multiple-qubit system in common dissipative environment \cite{JY}. In this paper, we are able to estimate entanglement dynamics by using the exact O-operators for the two-qubit models with common non-Markovian dissipative and dephasing environments, respectively.

\section*{Dissipation Model}
When $\mathcal{L}=\sigma^A_-+\kappa\sigma^B_-$, we can show that the O-operator takes the following form:
\begin{eqnarray}\non
\hat{\mathcal{O}}(t,s,x^*)&=&a(t,s)\sigma^A_-+b(t,s)\sigma^B_-+f(t,
s)\sigma^A_z\sigma^B_- \\ \label{Oop}
+g(t,s)\sigma^A_-\sigma^B_z&+&i\left(\int^t_0 ds'
p(t,s,s')x^*_{s'}\right)\sigma^A_-\sigma^B_-,
\end{eqnarray}
where equations of motion for $a(t,s)$, $b(t,s)$, $f(t,s)$, $g(t,s)$, and $p(t,s,s')$ can be derived from Eq. (\ref{consis2}). Then by definition, $\bar{\mathcal{O}}(t,x^*)=\int^t_0 ds \alpha(t,s)\hat{\mathcal{O}}(t,s,x^*)$
such that
\begin{eqnarray}\label{Obar}
\nonumber\\
\bar{\mathcal{O}}(t,
x^*)&=&A(t)\sigma^A_-+B(t)\sigma^B_-+F(t)\sigma^A_z\sigma^B_-
+G(t)\sigma^A_-\sigma^B_z\nonumber\\
&+&i\left(\int^t_0ds'P(t,s')x^*_{s'}\right)\sigma^A_-\sigma^B_-,
\end{eqnarray}
where $A(t)\equiv\int^t_0 ds \alpha(t,s)a(t,s)$, $B(t)\equiv\int^t_0 ds \alpha(t,s)b(t,s)$, $F(t)\equiv\int^t_0 ds \alpha(t,s)f(t,s)$, $G(t)\equiv\int^t_0 ds \alpha(t,s)g(t,s)$, and $P(t,s')\equiv\int^t_0 ds \alpha(t,s)p(t,s,s')$. \\

We can check that Eq.~(\ref{Oop}) indeed provides a consistent solution to Eq.~(\ref{consis2}). In fact, by substituting Eq.~(\ref{Oop}) into Eq.~(\ref{consis2}), the left-hand side (LHS) of it expands to
\begin{eqnarray*}
\frac{\partial\hat{\mathcal{O}}(t,s,x^*)}{\partial t}&=&
\frac{\partial a(t,s)}{\partial t}\sigma^A_-+\frac{\partial b(t,s)}{\partial t}\sigma^B_-+\frac{\partial f(t,s)}{\partial t}\sigma^A_z\sigma^B_-+\displaystyle\frac{\partial g(t,s)}{\partial t}\sigma^A_-\sigma^B_z \\ &+& ip(t,s,t)x^*_t+i\left(\int^t_0 ds'\frac{\partial p(t,s,s')}{\partial
t}x^*_{s'}\right)\sigma^A_-\sigma^B_-,
\end{eqnarray*}
while the right-hand side (RHS) of Eq. (\ref{consis2}) is composed of the following commutators:
\begin{eqnarray*}
\left[\frac{-i\omega_A}{2}\sigma^A_z,\hat{\mathcal{O}}(t,s,x^*)\right]
&=&i\omega_A[a(t,s)\sigma^A_- + g(t,s)\sigma^A_-\sigma^B_z\\
&+&i\left(\int^t_0 ds' p(t,s,s')x^*_{s'}\right)\sigma^A_-\sigma^B_-],
\end{eqnarray*}

\begin{eqnarray*}
\left[\frac{-i\omega_B}{2}\sigma^B_z,\hat{\mathcal{O}}(t,s,x^*)\right]
&=&i\omega_B[b(t,s)\sigma^B_- + f(t,s)\sigma^A_z\sigma^B_-\\
&+&i\left(\int^t_0 ds' p(t,s,s')x^*_{s'}\right)\sigma^A_-\sigma^B_-],
\end{eqnarray*}

\begin{eqnarray*}
[\mathcal{L}x^*_t,\hat{\mathcal{O}}(t,s,x^*)]=2x^*_t\left[f(t,s)+\kappa g(t,s)\right],
\end{eqnarray*}

\begin{eqnarray*}
&&-[\sigma^A_+\bar{\mathcal{O}}(t,x^*),\hat{\mathcal{O}}(t,s,x^*)]
=A(t)\left[a(t,s)\sigma^A_-+g(t,s)\sigma^A_-\sigma^B_z\right]\\
&&+iA(t)\left(\int^t_0
ds'p(t,s,s')x^*_{s'}\right)\sigma^A_-\sigma^B_-- B(t)\left[a(t,s)\sigma^A_z\sigma^B_-+g(t,s)\sigma^B_-\right]\\
&&+F(t)\left[a(t,s)\sigma^A_z\sigma^B_-+g(t,s)\sigma^B_-\right]+ G(t)[a(t,s)\sigma^A_-\sigma^B_z+g(t,s)\sigma^A_-]\\
&&+G(t)(b(t,s)+f(t,s))(\sigma^B_-+\sigma^A_z\sigma^B_-)+iG(t)\left(\int^t_0 ds'p(t,s,s')x^*_{s'}\right)\sigma^A_-\sigma^B_-]\\
&&-i\int^t_0 ds' P(t,s')x^*_{s'}[g(t,s)-a(t,s)]\sigma^A_-\sigma^B_-,
\end{eqnarray*}

\begin{eqnarray*}
&&-\kappa[\sigma^B_+\bar{\mathcal{O}}(t,x^*),\hat{\mathcal{O}}(t,s,x^*)]
=-\kappa A(t)[f(t,s)\sigma^A_-+b(t,s)\sigma^A_-\sigma^B_z]\\
&&+\kappa B(t)[b(t,s)\sigma^B_-+f(t,s)\sigma^A_z\sigma^B_-]+i\kappa B(t)\int^t_0 ds'p(t,s,s')x^*_{s'}\sigma^A_-\sigma^B_-\\
&&+\kappa F(t)[(a(t,s)+g(t,s))(\sigma^A_-+\sigma^A_-\sigma^B_z)]+\kappa F(t) [b(t,s)\sigma^A_z\sigma^B_-+f(t,s)\sigma^B_-]\\
&&+i\kappa F(t) \left(\int^t_0 ds'
p(t,s,s')x^*_{s'}\right)\sigma^A_-\sigma^B_-+\kappa G(t)[b(t,s)\sigma^A_-\sigma^B_z+f(t,s)\sigma^A_-]\\
&&i\kappa \left(\int^t_0 ds'
P(t,s')x^*_{s'}\right)[b(t,s)-f(t,s)]\sigma^A_-\sigma^B_-,
\end{eqnarray*}
as well as
\begin{eqnarray*}
-\mathcal{L}^{\dag}\frac{\delta
\bar{\mathcal{O}}(t,x^*)}{\delta
x^*_s}&=&-(\sigma^A_++\kappa\sigma^B_+)\frac{\delta[i\int^t_0 ds' P(t,s')x^*_{s'}]}{\delta x^*_s} \sigma^A_- \sigma^B_- \\
&=&-iP(t,s)\left[\frac{1}{2}(\sigma^A_z\sigma^B_- +
\sigma^B_-)+\frac{\kappa}{2}(\sigma^A_- + \sigma^A_-\sigma^B_z)\right].
\end{eqnarray*}
By equating the LHS with the RHS, we obtain the following partial differential equations for the coefficient functions $a(t,s), b(t,s), f(t,s), g(t,s)$ and $p(t,s,s')$:
\begin{eqnarray}\non
\frac{\partial a(t,s)}{\partial t}&=&i\omega_A a(t,s) + A(t)a(t,s) + G(t)g(t,s)+\kappa F(t) [a(t,s)+g(t,s)] \\\label{at}
&+&\kappa[G(t)-A(t)]f(t,s) -\frac{i\kappa}{2}P(t,s) ,
\end{eqnarray}

\begin{eqnarray}\non
\frac{\partial b(t,s)}{\partial t}&=&i\omega_B b(t,s) + \kappa B(t)
b(t,s)+\kappa F(t) f(t,s)\\ \label{bt}&+&G(t)[b(t,s)+f(t,s)] +[F(t)-B(t)]g(t,s)-\frac{i}{2}P(t,s),
\end{eqnarray}

\begin{eqnarray}\non
\frac{\partial f(t,s)}{\partial t}&=&i\omega_B f(t,s) +\kappa F(t) b(t,s)+\kappa B(t) f(t,s)\\ \label{ft} &+&G(t)[b(t,s)+f(t,s)]+[F(t)-B(t)]a(t,s)-\frac{i}{2}P(t,s),
\end{eqnarray}

\begin{eqnarray}\non
\frac{\partial g(t,s)}{\partial t}&=&i\omega_A g(t,s) + G(t) a(t,s) +A(t)g(t,s)+\kappa F(t)[a(t,s)+g(t,s)]\label{gt}\\ &+& \kappa[G(t)-A(t)]b(t,s)-\frac{i\kappa}{2}P(t,s),
\end{eqnarray}

\begin{eqnarray}\non
\frac{\partial p(t,s,s')}{\partial t}&=&i(\omega_A+\omega_B)p(t,s,s')
+[A(t)+G(t)+\kappa B(t) + \kappa F(t)] p(t,s,s') \\ \label{pt}
&+&P(t,s')[a(t,s)-g(t,s)+\kappa b(t,s)-\kappa f(t,s)],
\end{eqnarray}
as well as the boundary condition
\begin{eqnarray}\label{pbc}
p(t,s,t)=-2if(t,s)-2i\kappa g(t,s).
\end{eqnarray}
We also deduce the initial conditions $a(s,s)=1$, $b(s,s)=\kappa$, and $f(s,s)=g(s,s)=p(s,s,s')=0$ from the fact that $\hat{\mathcal{O}}(s,s,x^*)=\mathcal{L}$.

Eqs.~(\ref{at}-\ref{pt}) are the required exact equations that govern the O-operator evolution and, in principle, allow us to numerically solve the QSD equation. We now consider the bath correlation function to be an Ornstein-Uhlenbeck process such that $\alpha(t,s)=\frac{\gamma}{2}e^{-\gamma|t-s|}$, which facilitates a set of simpler ordinary differential equations from the above Eqs.~(\ref{at}-\ref{pt}), as presented in Eq.~(\ref{ABFGQ}). For instance, we have
\begin{eqnarray}\non
\frac{\partial A(t)}{\partial t}&=&\frac{\partial}{\partial
t}\int_0^tds\alpha(t,s)a(t,s)\\  &=&\alpha(t,t)a(t,t)+\int_0^tds\frac{\partial\alpha(t,s)}{\partial
t}a(t,s)+\int_0^tds\alpha(t,s)\frac{\partial a(t,s)}{dt}\non \\
&=& \frac{\gamma}{2}-\gamma\int_0^tds\alpha(t,s)a(t,s)+\int_0^tds\alpha(t,s)[i\omega_A a(t,s)+A(t)a(t,s)\non\\ &+& G(t) g(t,s)+ \kappa F(t)(a(t,s)+g(t,s))+\kappa(G(t)-A(t))f(t,s)-\frac{i\kappa}{2}P(t,s)] \non\\
&=& \frac{\gamma}{2}-\gamma A(t)+i\omega_AA(t)+A^2(t)+G^2(t)+2\kappa F(t)G(t)-\frac{i\kappa}{2}Q(t) \label{At},
\end{eqnarray}
where $Q(t)\equiv\int_0^tds\alpha(t,s)P(t,s)$. It is noted that the initial condition $a(s,s)=1$ has been used. Applying a similar derivation, we also have
\begin{eqnarray}\label{Bt}
\frac{\partial B(t)}{\partial t}&=&\frac{\gamma\kappa}{2}-\gamma
B(t)+i\omega_BB(t)+\kappa B^2(t)+\kappa F^2(t)\non\\
&+&2G(t)F(t)-\frac{i}{2}Q(t),
\end{eqnarray}
\begin{eqnarray}\label{Ft}
\frac{\partial F(t)}{\partial t}&=&-\gamma F(t)+i\omega_B F(t)+2\kappa B(t)F(t)+B(t)G(t)\non\\
&+&F(t)G(t)+A(t)F(t)-A(t)B(t)-\frac{i}{2}Q(t),
\end{eqnarray}
\begin{eqnarray}\label{Gt}
\frac{\partial G(t)}{\partial t}&=&-\gamma G(t)+i\omega_A G(t)+2A(t)G(t)+\kappa A(t)F(t)\non\\
&+&\kappa F(t)G(t)+\kappa B(t)G(t)-\kappa A(t)B(t)-\frac{i\kappa}{2}Q(t),
\end{eqnarray}
and
\begin{eqnarray}\label{Pts}
\frac{\partial P(t,s')}{\partial t}=-\gamma
P(t,s')+i(\omega_A+\omega_B)P(t,s')+[2A(t)+2\kappa B(t)]P(t,s').
\end{eqnarray}
By the boundary condition in Eq.~(\ref{pbc}) and the definitions of $P(t,s')$, $F(t)$, and $G(t)$, it is found $P(t,t)=-2iF(t)-2i\kappa G(t)$ and the solution of Eq.~(\ref{Pts}) is Eq.~(\ref{sPts}).

To construct a closed group of differential equations for Eqs.~(\ref{At}-\ref{Gt}), we derive the ordinary differential equation for $Q(t)$ using Eq.~(\ref{Pts}):
\begin{eqnarray}\non
\frac{\partial Q(t)}{\partial t}&=&\frac{\partial}{\partial
t}\int_0^tds'\alpha(t,s')P(t,s')\\ &=&\alpha(t,t)P(t,t)+\int_0^tds'\frac{\partial\alpha(t,s')}{\partial
t}P(t,s')+\int_0^tds'\alpha(t,s)\frac{\partial P(t,s')}{dt} \non\\
&=&\frac{\gamma}{2}[-2iF(t)-2i\kappa
G(t)]-\gamma\int_0^tds'\alpha(t,s')P(t,s')\non\\
&+&\int_0^tds'\alpha(t,s')[-\gamma P(t,s') +i(\omega_A+\omega_B)P(t,s')\non\\
&+&(2A(t)+2\kappa B(t))P(t,s')] \non\\
&=&-i\gamma[F(t)+\kappa G(t)]-2\gamma Q(t)+i(\omega_A+\omega_B)Q(t)\non\\
&+&[2A(t)+2\kappa B(t)]Q(t)\label{Qt} .
\end{eqnarray}
By definition, the initial conditions for the coefficients of $\bar{\mathcal{O}}(t,x^*)$ are  $A(0)=B(0)=F(0)=G(0)=Q(0)=0$.

\end{document}